\documentclass[aps,pre,twocolumn,showpacs,
amsmath,amssymb,reprint]{revtex4}

\usepackage{graphicx}
\usepackage{dcolumn}
\usepackage{bm}
\begin{document}

\preprint{APS/123-QED}

\title{Diffusivity, mobility and time-of-flight measurements:\\
scientific fictions  and 1/f-noise realization }

\author{Yuriy E. Kuzovlev}
\affiliation{Donetsk A.A.Galkin Institute
for Physics and Technology of NASU,\\
ul.\,R.\,Luxemburg 72, Donetsk 83114, Ukraine}
\email{kuzovlev@fti.dn.ua}

\date{27 August 2010}

\begin{abstract}
Influence of equilibrium thermal 1/f-type mobility fluctuations on
time-of-flight measurements is considered. We show that it can
explain experimental time dependencies of transient photocurrents.
\end{abstract}



\maketitle


\section{\,Introduction}

It is well known that definite aspects of chaotic motion of
quasi-free charge carriers in many materials can be described in
terms of diffusivity, $\,D\,$, and mobility, $\,\mu\,$. According to
definitions of these quantities in statistical physics, if\,
$\,\Delta X(t)\,$\, is displacement, or path, of a carrier during
time interval $\,(0,t)\,$, and $\,t\,$ much exceeds characteristic
memory time, $\,\tau_0\,$, of carrier's random walk, then
\begin{eqnarray}
\frac 1t\, \left\langle\,\Delta
X^2(t)\,\right\rangle_0\,=\,2D\,\,\,,\label{d}\\
\frac 1t\, \langle\,\Delta X(t)\,\rangle_f\,=\,
\mu\,f\,\,\,,\label{mu}
\end{eqnarray}
where the angle brackets $\,\langle ...\rangle_0\,$  and $\,\langle
...\rangle_f\,$ denote averaging over thermodynamically equilibrium
statistical ensemble and  non-equilibrium one in presence of
sufficiently weak force $\,f\,$ per carrier (to be precise, we
consider projections of vectors just onto direction of this force).

Thus, $\,D\,$ and $\,\mu\,$ are averages over infinite variety of
carriers. However, in practice many authors interpret relations
(\ref{d})-(\ref{mu}) in the form like
\begin{eqnarray}
\Delta X(t)\,=\, \mu ft\,+\,\Delta_d X(t)\,\,\,,\label{si}\\
\left\langle\,\Delta_d
X^2(t)\,\right\rangle_f\,=\,2Dt\,\,\,,\nonumber
\end{eqnarray}
where $\,\Delta_d X(t)\,$ is ``diffusive component'' of the
displacement. Thus, the ensemble averages\,$\,D\,$ and $\,\mu\,$ are
assigned to a separate carrier as characteristics of its individual
random trajectory!

Such the arbitrariness, or ``scientific fiction'', implies illusion
of complete knowledge of non-equilibrium noise which stays the same
as in equilibrium. Consequently, in view of the Einstein relation,
$\,D=T\mu\,$, one may think that at\, $\,\phi(t) \gg 1\,$\,, where
\[\phi(t)\,\equiv\,\frac {f\sqrt{Dt}}T\,\,\,, \]
the carriers' displacement is almost non-random, and therefore the
time-of-flight measurements of transient photocurrents (see e.g.
works \cite{uch,nir,reviz,sil,nd} and references therein) should give
nearly ideal responses to pulse-like or step-like excitation (see
Fig.1 and Fig.2).

In reality  one observes \cite{uch,nir,reviz,sil,nd} something
qualitatively different, similar to what is shown in in Fig.1 and
Fig.2.

Such the difference is conventionally explained as result of random
variations of time of flight (TOF) because of trapping and
de-trapping of excess carriers by localized states (the so-called
``dispersive transport'', see e.g. \cite{uch,zv} and references
therein). This explanation indeed is good in many characteristic
cases. But in many other cases it seems unsatisfactory. Perhaps, by
this reason, for instance, in \cite{nir} the idea of ``frozen''
spatial variations of mobility was suggested.

Quite another explanation arises when one understand that the
fantastic expression (\ref{si}) should be replaced by a realistic
one,
\begin{eqnarray}
\Delta X(t)\,=\, \widetilde{\mu}(t) ft\,+\,\Delta_d X(t)
\,\,\,,\label{ri}\\
\left\langle\,\widetilde{\mu}(t)\,\right\rangle_f\,=\,\mu\,\,\,,\nonumber
\end{eqnarray}
where\, $\,\widetilde{\mu}(t)\,$\, is random quantity even in absence
of the trapping, merely because a separate carrier has no definite
mobility (as well as a separate person does not have ``mean life''
and even may not know about it)!

This statement is irrefutable, but its comprehension is prevented by
prejudices of the naive probability-theoretical way of thinking
instead of the fundamental statistical-mechanical one \cite{kr}. For
the first this statement was argued in
\cite{pr1,pjtf,bk1,bk2,bk3,pr2} in the course of unprejudiced
statistical analysis of random walks of atomic-size particles and
later in \cite{i1} (or see \cite{bbgky}) and in \cite{p12,feb,tmf,p4}
was confirmed, by example of fluid molecules, as consequence of the
first-principle equations of statistical mechanics.

Importantly, according to these investigations, a magnitude
 of fluctuations (uncertainty \cite{bk1}) of
individual mobility of charge carrier is generally on order of it
itself (i.e. $\,\widetilde{\mu}-\mu\sim\mu\,$). Therefore probability
distributions of\, $\,\Delta X(t)\,$ taken at different values of the
dimensionless force\, $\,\phi=\phi(t)\,$ (see above) differ one from
another not only by their mean positions but also by their widths and
shapes, as it is shown in Fig.3, so that real drift of a packet of
even non-interacting carriers is accompanied by its spreading,
approximately proportionally to both $\,t\,$ and $\,f\,$. Particular
results of such drift are the dark-curve time dependencies in
Figs.1-2.

The purpose of the present paper is just to demonstrate that these
effects of the native carrier's mobility fluctuations resemble
effects of the trapping.

\section{Why and how mobility fluctuates by itself}

Introducing related fluctuating diffusivity
$\,\widetilde{D}(t)=T\widetilde{\mu}(t)\,$, one can ask equivalently:
why carrier's diffusivity fluctuates by itself? The cause was
explained in the mentioned works. It is temporal non-locality of such
physical characteristics as diffusivity: the latter by its sense
describes random motion of a carrier (or a particle in general)
during time intervals significantly greater than the memory time
$\,\tau_0\,$. Therefore at the end of any such interval the system
already does not remember what took place at its beginning.
Consequently, the system can not keep a definite value of the
diffusivity.

\begin{figure}
\begin{center}
\includegraphics[width=8cm,height=7cm]{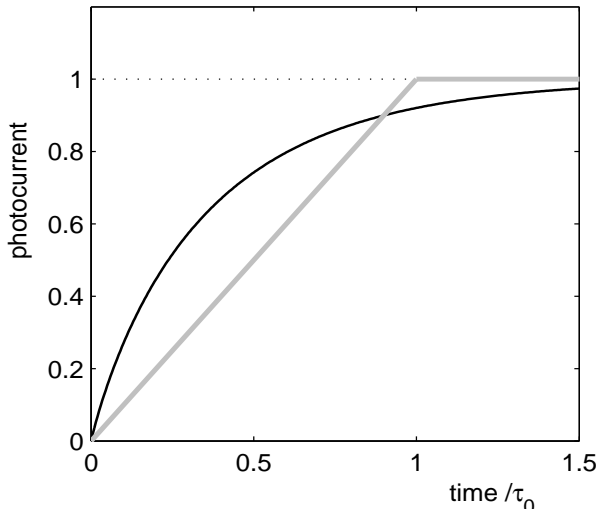}
\end{center}
\caption{\label{F1} \,Ideal (fat grey curve) and actual (dark curve)
transient photocurrent responses to switching on illumination. The
quantity $\tau_o\,$ is is mean time of flight. }
\end{figure}

If it is seen, then it is clear that resulting fluctuations
(uncertainty) of diffusivity (and hence mobility), firstly, by their
order of magnitude  are comparable with its average value. Secondly,
they can be arbitrary long, that is have no characteristic  time
scale, and thus represent a ``flicker noise'' whose power spectrum
$\,S_D(\omega)\,$ unboundedly grows at $\,\omega\rightarrow 0\,$.
Thirdly, their probability distribution is close to a stable one.

\begin{figure}
\begin{center}
\includegraphics[width=8cm,height=7cm]{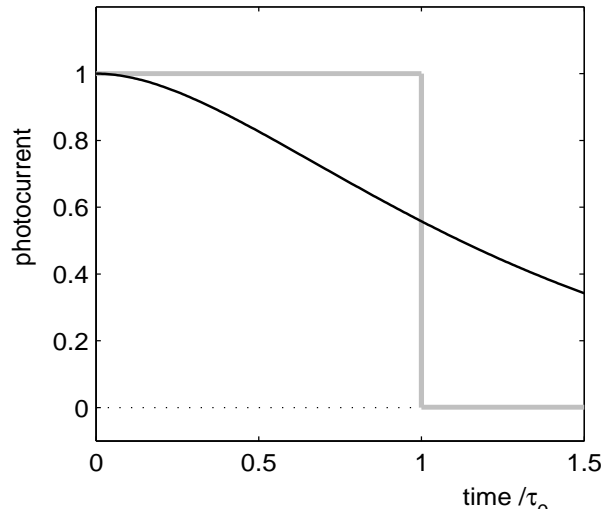}
\end{center}
\caption{Ideal (fat grey curve) and actual (dark curve) transient
photocurrent responses to short pulse of illumination.  }
\end{figure}

\begin{figure}
\begin{center}
\includegraphics[width=8cm,height=7cm]{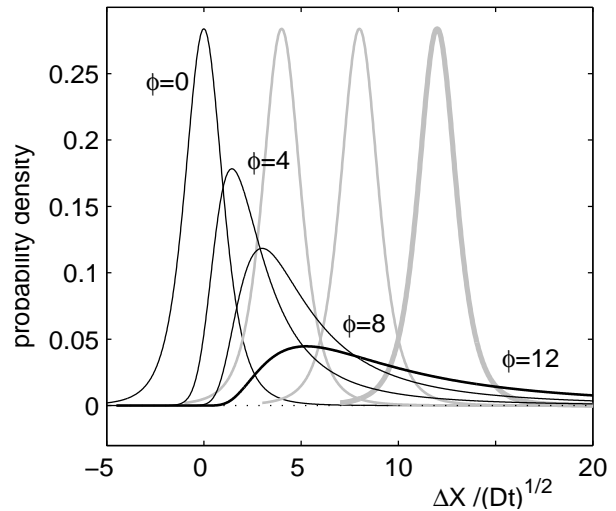}
\end{center}
\caption{ Comparison of evolutions of the $\,\Delta X(t)\,$
distribution, considered as a function of the dimensionless parameter
$\,\phi\,$, in the conventional ``fantastic'' model (\ref{si}) (four
identical symmetric bell--shaped curves) and in realistic model
(\ref{ri}) according to results of \cite{bk1,bk2,pr2} (dark
asymmetrical curves). The most left bell--shaped curve corresponding
to $\,\phi=0\,$ is common for both models. The two most right fat
curves demonstrate qualitative difference of the models at
$\,\phi=12\,$. }
\end{figure}

Additional reasonings can be found in
\cite{bk1,bk2,bk3,i1,bbgky,i2,p12,feb,tmf,p4,i3,i4,thy1,thy2}. To
confirm them at fundamental level, we have to calculate statistical
characteristics of the displacement $\,\Delta X(t)\,$, such as its
characteristic function $\,\langle \exp{[ik\Delta X(t)]}\rangle\,$,
merely following rigorous recipes and equations of the statistical
mechanics. Or, at least, analyze admissible forms of $\,\langle
\exp{[ik\Delta X(t)]}\rangle\,$ at phenomenologically under minimum
of assumptions.

\begin{figure}
\begin{center}
\includegraphics[width=8cm,height=7cm]{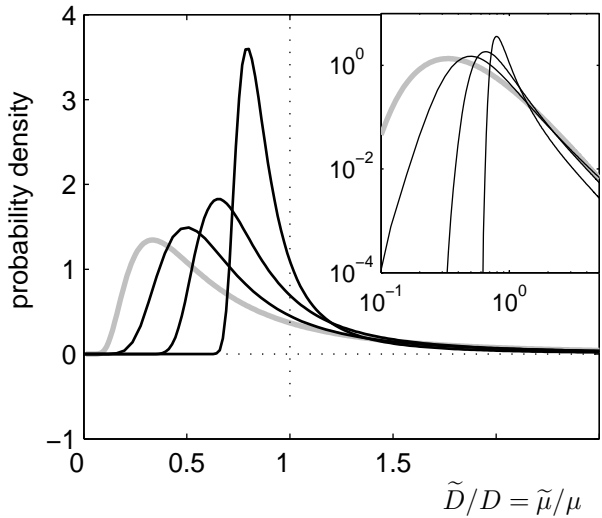}
\end{center}
\caption{The three dark curves represents the diffusivity and
mobility probability density distributions following from Eq.\ref{cf}
\cite{bk2,pr2,i2} at different values of $\,\ln{(t/\tau_0)}\,$ and
$\,r^2_0/D\tau_0\,$. The fat grey curve illustrates the diffusivity
and mobility distribution (10) \cite{p12,feb,tmf,p4}. All three
distributions have power-law long tail with the exponent between 2
and 3 (somehow cut off far on the right). The inset: the same in
double-logarithmic scale. }
\end{figure}

For the first such investigation was made in \cite{pr1,pjtf,bk1} (or
see \cite{i2}) where thermodynamically equilibrium random walk
$\,\Delta X(t)\,$ of an electron (or other particle) in spatially
uniform medium was considered basing on the only assumption that it
possesses diffusive scale invariance, $\,\Delta X(st)\sim
\sqrt{s}\,\Delta X(t)\,$ (the tilde symbolizes statistical
equivalence) but destroyed at some microscopic time scales
$\,\tau_0\,$ and $\,r_0\,$ (e.g. mean free path or mean hop length of
the particle under consideration). It was shown that at $\,t \gg
\tau_0\,$ and $\,|k|r_0\ll 1\,$ the characteristic function of
$\,\Delta X(t)\,$ can be approximately expressed by
\begin{eqnarray}
\langle \exp{[ik\Delta X(t)]}\rangle\,=\,
\exp{\left[\,-Dk^2t\,\,\frac
{\ln{(\tau_0/t+r_0^2k^2)}}{\ln{(\tau_0/t)}}\right]} \,\,\,
\label{cf0}
\end{eqnarray}
The corresponding statistics can be named ``quasi-Gaussian''
\cite{pr2,i2}.

Similar analysis of non-equilibrium situation, when the particle is
affected by an external force $\,f\,$, needs in the help of the
``generalized fluctuation-dissipation relations'' \cite{fds,p} which
show \cite{bk2,bk3,pr2,i1,i2,feb} that at sufficiently weak force, in
 the sense of $\,fr_0/T \ll 1\,$, the characteristic function can be
 obtained by simple replacement in (\ref{cf0}), namely,
 \begin{eqnarray}
-k^2\, \rightarrow\, -k^2+ikf/T  \,\,\, \label{ch}
\end{eqnarray}
(thus the Einstein relation extends from mean values to fluctuations
of diffusivity and mobility).

\begin{figure}
\begin{center}
\includegraphics[width=8cm,height=7cm]{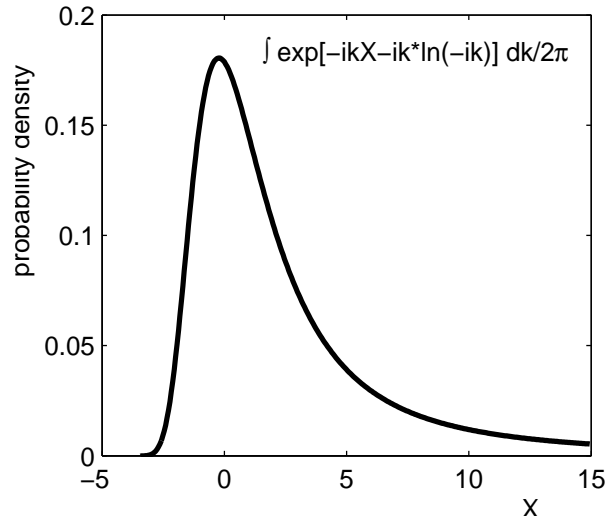}
\end{center}
\caption{The maximally asymmetric stable Cauchy distribution. It is
principal ancestor of the distributions shown in Figs.3 and 4.}
\end{figure}

Therefore, at $\,\phi(t) \gg 1\,$, neglecting in (\ref{ri}) the
diffusive part $\,\Delta_dX(t)\,$ of $\,\Delta X(t)\,$, one can write
approximately
\begin{eqnarray}
\langle \exp{[ik\Delta X(t)]}\rangle\,=\, \exp{\left[\,ik\mu f
t\,\,\frac {\ln{(\tau_0/t-ikfr_0^2/T)}}{\ln{(\tau_0/t)}}\right]}
\,\,\, \label{cf}
\end{eqnarray}

The corresponding statistics can be named ``quasi-Cauchy'' since the
corresponding probability distributions (see dark curves in  Fig.4)
are similar to the asymmetric stable Cauchy distribution
\cite{fel,luc} (see Fig.5).

Just such statistics implies the displacement distributions shown in
Fig.3 by dark curves. Its physical meaning was explained in
\cite{bk2,pr2,i2}.

One can see that characteristic functions like (\ref{cf0}) and
(\ref{cf}) mean presence of ``flicker'', or 1/f-noise type,
fluctuations in diffusivity and mobility, with low-frequency power
spectrum \cite{pjtf,bk1,bk2,bk3,i2}
\begin{eqnarray}
\frac {S_D}{D^2}= \frac {S_{\mu}}{\mu^2}= \frac
{2\pi\alpha(\omega)}{\omega}\,\,\,,\,\,\,\,\, %
\alpha(\omega)\approx \frac {2r_0^2}{3D\tau_0\,\ln^2(\tau_0\omega)}
\,\, \label{sp}
\end{eqnarray}
This theoretical estimate very well agreed with experimental data in
1983 \cite{bk3} and equally well agrees with new data about
semiconductors \cite{sem}, carbon nano-tubes \cite{nt}, graphene
layers and graphene transistors \cite{av,balan} and other electronic
devices (at that revealing origin of typical values of the ``Hooge
constant'', $\,\alpha(\omega)\sim 0.002\,$ \cite{bk3}). It should be
underlined again and again that this 1/f-noise is by its nature an
integral part of the atomic thermal motion and exists even in ideal
crystal structure in absence any traps (see explanations in
\cite{bk1,bk3,i2,i3,i4})!

The fundamental microscopic approach to analogous diffusivity and
mobility fluctuations so far was tried only in case of molecular
random walks in fluids \cite{i1,bbgky,i2,p12,tmf,feb,p4}, on the base
of the exact Bogolyubov equations \cite{bog}.

In \cite{i1} (see also \cite{i2,bbgky}) it was found, for atoms of a
gas, that
\begin{eqnarray}
\frac {S_D(\omega)}{D^2}= \frac {S_{\mu}(\omega)}{\mu^2}\approx \frac
{\pi }{\omega}\,\ln{\frac 1{\tau_0\omega}}\,\,\,,\, \label{sp1}
\end{eqnarray}
which satisfactorily agrees with experimental data on electrolytes
\cite{bk3,b} (where the Hooge constant is anomalously large).

\begin{widetext}

\begin{figure}
\begin{center}
\includegraphics[width=17cm,height=7cm]{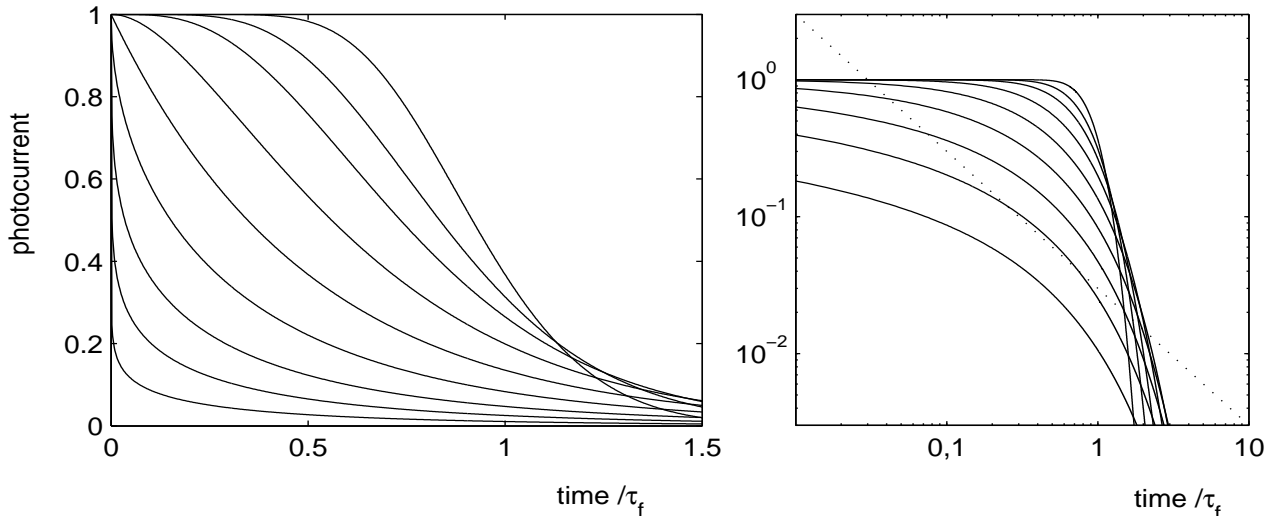}
\end{center}
\caption{On the left:\, transient photocurrent response to short
pulse illumination corresponding to the mobility distribution (10) at
different $\,\beta\,$ values. On the right:\, the same in double
logarithmic scale. The dot line indicates slope -1.}
\end{figure}

\end{widetext}

\begin{figure}
\begin{center}
\includegraphics[width=8cm,height=7cm]{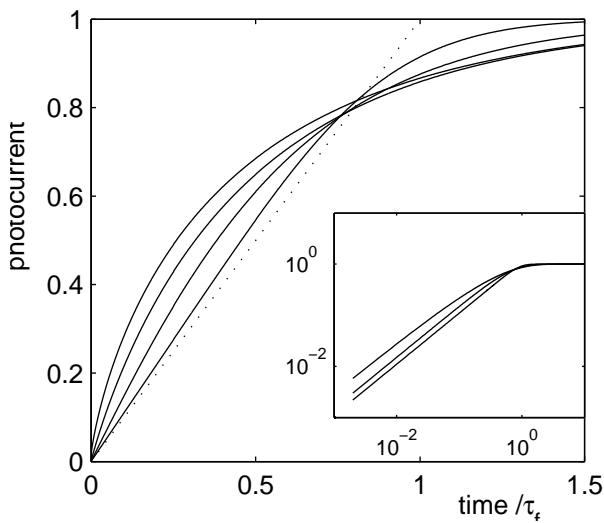}
\end{center}
\caption{Transient photocurrent response to step pulse illumination
corresponding to the mobility distribution (10) at different
$\,\beta\,$ values. In the inset:\, the same in double logarithmic
scale.}
\end{figure}

But the underlying probability distribution of diffusivity and
mobility remained unknown. It was found in \cite{p12,p4} by means of
more rough but deeper penetrating approximation, and its density can
be expressed as
\begin{eqnarray}
U_\beta(x)= \frac
1{x^{\,\beta+2}\,\Gamma(\beta+1)}\,\exp{\left(-\frac 1x
\right)}\,\,\,, \label{gas}\\
\,\,\,x= \frac {\widetilde{D}}{\beta D}=\frac
{\widetilde{\mu}}{\beta\mu}\,\, \,,\nonumber
\end{eqnarray}
with $\,\beta>0\,$. The spectrum (\ref{sp1}) arose in the case when
$\,\beta=1\,$ \cite{p12}. The corresponding probability density
$\,U_1(x)\,$ is shown in Fig.4 by far grey curve. Evident property of
all obtained distributions is that their most probable values are
significantly smaller (three times at $\,\beta=1\,$) than their mean
values.

Thus, both phenomenological and microscopical approaches to the
native thermal mobility fluctuations (when being correctly performed)
gave qualitatively identical results.

\section{Manifestation of the native 1/f mobility fluctuations in
TOF experiments}

The TOF measurements of excess (injected) probe particles, e.g.
photo-excited charge carriers, usually presume that $\,\phi(t)\gg
1\,$. According to the aforesaid, this means that $\,S_{\mu}f^2
\gtrsim 2T\mu\,$, that is at corresponding time intervals
 contribution to particle's path fluctuations from 1/f
diffusivity/mobility fluctuations exceeds contribution from diffusion
itself (i.e. from white thermal noise). In other words, TOF
experiments are first of all observations of the native
diffusivity/mobility 1/f-noise.

Hence, it is reasonable first to consider such imaginary (but not
fantastic) situations when the above mentioned simple theory is
sufficiently adequate. This would determine a background of the TOF
spectra which exists
 even in absence of specific effects like the ``dispersive
 transport''  (transport via a large number of free electron traps)
 jot taken into account by the theory.

At $\,\phi(t)\gg 1\,$, if knowing the mobility probability
distribution, one can easy find that of TOF $\,\tau=L/\mu f\,$ (where
$\,L\,$ is distance to be overcame) and resulting time dependence of
transient current (flow), and vice versa. The distributions
corresponding to characteristic function (\ref{cf}) have no simple
analytic expression but can be approximated by simple related
distributions (10). Then calculation of of transient current become
so trivial that its description can be omitted.

The results for short-pulse and step-like excitations are
demonstrated on Fig.6 and Fig.7 where the currents are normalized to
unit start value and unit steady-state value, respectively
($\,\tau_f\,$ there is mean TOF value).

\section{Discussion and conclusion}

Visual comparison of our Figs.6 and 7 with experimental data
graphically resented in \cite{uch,nir,reviz,sil,nd,zv} (and similar
works) shows rather high degree of correlation between one and
another. This, gives evidence that many observations can be explained
by equilibrium mobility/diffusivity 1/f-type fluctuation accompanying
usual, ``non-dispersive'', transport. On the other hand, we can say
that many observations, when even their authors have doubts as to
``dispersive transport'', do proof that the latter is not unique
cause of TOF and mobility distributions.

For instance, may be, the experimental mobility distribution found in
work \cite{nir} (which stimulated the present work) should be
assigned not to spatial inhomogeneities but to temporal mobility
fluctuations of charge carriers (otherwise, why this distribution so
strongly resembles the picture in Fig.4 ?). Analogously, the large
width and asymmetry of mobility distributions of ions in air measured
in \cite{dma} can be manifestation of flicker fluctuations  in ions'
mobilities.

At the same time, undoubtedly, our theoretical gallery does not
include specific pictures of pronounced ``dispersive transport''.
This means merely that  1/f-noise under dispersive transport has more
rich scaling properties and hence more complicated mobility
statistics than what was considered; Developing of theory of such
1/f-noise from rigorous statistical mechanics, instead of
probabilistic ``scientific fiction'', is interesting task for the
future.


\end{document}